\documentclass[pra,twocolumn,superscriptaddress,showpacs,amsmath,amssymb,longbibliography,aps,10pt]{revtex4-2}
\usepackage{graphicx, color}
\usepackage{dcolumn}
\usepackage{bm}
\usepackage{epstopdf}
\usepackage{physics}
\usepackage[dvipsnames]{xcolor}
\usepackage{subfigure}
\usepackage{pifont}

\definecolor{orange}{rgb}{1,0.5,0}
\definecolor{goodgreen}{rgb}{0.1,0.5,0}
\definecolor{goodred}{rgb}{0.7,0,0}
\usepackage[colorlinks,urlcolor=goodgreen,citecolor=blue,linkcolor=goodred]{hyperref}
\usepackage[normalem]{ulem}
\usepackage{verbatim}
\usepackage{dsfont}

\graphicspath{ {./images/} }

\let\oldepsilon\epsilon \let\epsilon\varepsilon \let\varepsilon\oldepsilon

\begin{document}
\title{Squeezing and measurement of a mechanical quadrature via PID feedback} 

\begin{abstract}
Proportional-Integral-Derivative (PID) control is used for automatically regulating a measurable quantity to a desired setpoint. It is widely used in different types of classical control electronics. Here, we show how generalizing the feedback term in quantum systems to include derivative and integral components influences both the transient and steady-state behavior of the amplitude and squeezing of a mechanical quadrature in an optomechanical system. We show that, the derivative feedback leads to enhanced unconditional squeezing. Furthermore, we demonstrate how feedback may be employed to drive a mechanical quadrature to a desired setpoint, and to track an external force. Our findings offer new routes for an improved quantum state control and measurement precision.
\end{abstract}

\author{Alberto Hijano}
\email{alhijano@jyu.fi}
\affiliation{Department of Physics and Nanoscience Center, University of Jyväskylä, P.O. Box 35 (YFL), FI-40014 University of Jyväskylä, Finland}

\author{Tero T. Heikkil\"a}
\email{tero.t.heikkila@jyu.fi}
\affiliation{Department of Physics and Nanoscience Center, University of Jyväskylä, P.O. Box 35 (YFL), FI-40014 University of Jyväskylä, Finland}

\maketitle

\section{Introduction}
\label{sec:introduction}

Quantum systems exhibit intrinsic measurement uncertainty arising from quantum fluctuations. Furthermore, the measurement process itself introduces back action, which can fundamentally limit the achievable measurement precision. If an observable commutes with the system Hamiltonian, its uncertainty does not increase under unitary time evolution. This principle is the basis of non-demolition measurements~\cite{Braginsky:1996}. Several quantum non-demolition measurements have been proposed for the detection of weak external forces~\cite{Braginsky:1980,Kimble:2001,Purdue:2002}. Deviations from the predicted evolution of the monitored observable provide information about the forces acting on the system. While the measurement of an observable may induce conditional squeezing, i.e. the system is squeezed towards the state corresponding to the measurement outcome in any given run of the experiment, excess noise arising from ensemble averaging masks such squeezing, so it is not sufficient to induce unconditional squeezing affecting the variance over all measurement outcomes. Unconditional squeezing may be achieved via measurement-based negative feedback~\cite{Wiseman:1994}, using resolved sideband cooling with squeezed or modulated input light~\cite{Jahne:2009,Mari:2009}, or by using coherent quantum feedback~\cite{Bemani:2024,Jiao:2026}.

Braginsky \textit{et al.} proposed a back-action evasion measurement scheme for a mechanical quadrature~\cite{Braginsky:1980,Braginsky-book,Malz:2016}. In that scheme, the mechanical resonator is parametrically coupled to an optical cavity, which is driven by an amplitude-modulated field. The quadrature can be detected via a homodyne measurement of the signal leaving the cavity. Such measurement may be used to induce squeezing of the quadrature by applying a force to the resonator proportional to the measurement~\cite{Clerk:2008}. More recently, Di Giovanni \textit{et al.}~\cite{DiGiovanni:2021} proposed to use Bayesian linear–quadratic–Gaussian feedback to achieve optimal unconditional stationary squeezing. Other types of measurement-based feedback strategies have also been proposed to achieve mechanical cooling and squeezing, such as the displacement of the input field~\cite{Isaksen:2023}, and the amplitude modulation of the input light~\cite{Wang:2026}.

While back-action evasion measurements have been carried out for the detection of weak forces~\cite{Marchese:1992,Bocko:1996}, there are no experimental realizations of measurement-based feedback-induced squeezing. The main challenge is processing the measurement and applying the feedback coherently. We believe this will be realized in the near future, as ground state feedback cooling has already been achieved in mechanical oscillators~\cite{Rossi:2018,Whittle:2021,Rej:2025}. Moreover, this scheme is constrained by the requirement of a strong coupling to the optical mode, the requirement of a low mechanical occupation and the side-effect of parametric instability due to strong radiation pressure~\cite{Hertzberg:2010}.

\begin{figure}[t!]
    \centering
    \includegraphics[width=0.99\columnwidth]{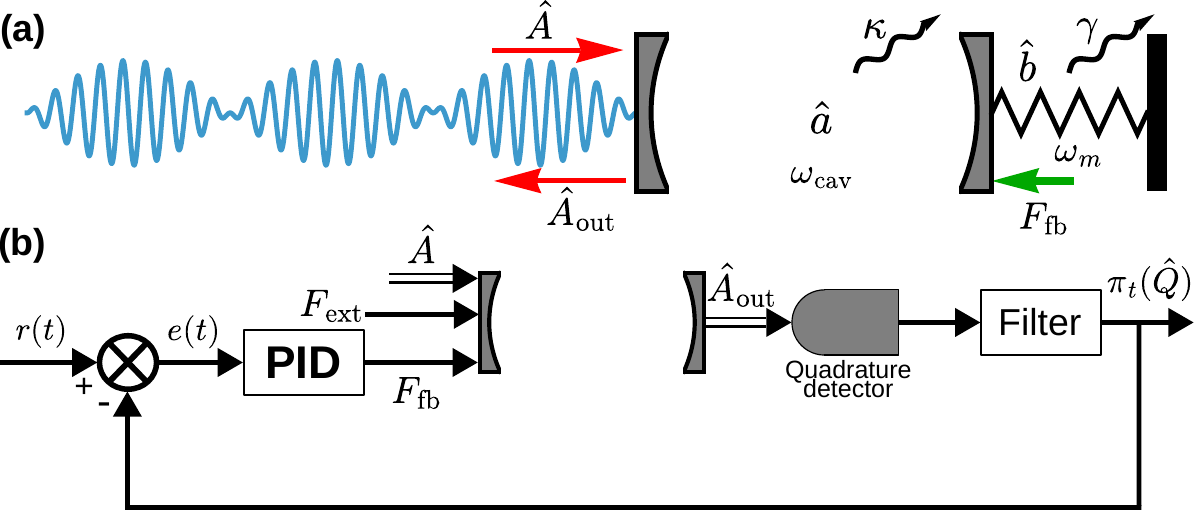}
    \caption{(a) Sketch of the optomechanical system. The cavity is driven by an amplitude-modulated input field. The output of the detector, which is coupled to the $\hat{Q}$ quadrature, is used to apply a feedback force $F_\mathrm{fb}$. (b) Diagram of the PID-controlled feedback loop. Double-lined arrows denote quantum (operator-valued) signals. The feedback Hamiltonian $\hat{H}_\mathrm{PID}$ depends on the PID controlled error signal $e(t)=r(t)-\pi_t(\hat{Q})$, where $r(t)$ is the setpoint and $\pi_t(\hat{Q})$ is the estimated value of $\hat{Q}(t)$.}
    \label{fig:sketch}
\end{figure}

In this work, we generalize the proportional feedback force to account for integral and derivative control (PID control). PIDs have extensively been used in the control of classical systems, so there is a large body of knowledge about their underlying theory~\cite{Ziegler-Nichols,Borase:2021}. By this example, we show how they can also be used for practical applications in quantum control and quantum technologies, for instance for enhancing the accuracy of the measurement of a time-dependent force. For this purpose, we consider a quantum Kalman filter based PID controller, developed in Ref.~\cite{Gough:2020}.

We show that PID control enhances the transient and steady-state response of the squeezing of a mechanical quadrature in an optomechanical system, a purely quantum effect with no classical counterpart. Moreover, we apply the PID control to drive the mechanical quadrature to the desired setpoint. Finally, we show how PID feedback may be used to speed up the tracking of an external force in a sensing experiment. We use the SLH formalism~\cite{Gough:networks,Gough:series-product,Combes:2017}, a synthesis of the quantum stochastic calculus developed by Hudson-Parthasarathy in 1984~\cite{Hudson:1984,Parthasarathy-book} and the quantum input-output theory developed by Gardiner and Collett~\cite{Gardiner-Collett-input-output,Zoller-Gardiner-book}. We derive full analytical equations for the quadrature uncertainties, and show that the derivative action may be used to speed up and enhance the unconditional squeezing. We show that tuning the PID parameters allows for precise control of the quadrature amplitude, for instance, for the initialization of the resonator in a sensing experiment.

\section{Model and formalism}
\subsection{Filtering of an optomechanical system}
A mechanical oscillator parametrically coupled to an electromagnetic cavity is described by the Hamiltonian~\cite{Aspelmeyer:2014}:
\begin{equation}
    \hat{H}_\mathrm{sys}=\hbar\omega_\mathrm{cav}\hat{a}^\dagger\hat{a}+\hbar\omega_m\hat{b}^\dagger\hat{b}-\hbar G_0(\hat{b}^\dagger+\hat{b})\hat{a}^\dagger\hat{a}\; ,
\end{equation}
where $\hat{a}$ and $\hat{b}$ are the cavity and mechanical resonator annihilation operators, $\omega_\mathrm{cav}$ is the cavity resonance frequency, $\omega_m$ is the mechanical resonance frequency and $G_0$ is the optomechanical coupling strength.

Single quadrature detection may be accomplished by modulating the optomechanical coupling at the oscillator frequency~\cite{Caves:1980,Bocko:1996}. The mechanical motion is correlated with the output signal from the cavity, so one may perform a homodyne measurement of the output signal to perform a non-demolition measurement of a mechanical quadrature. We follow the double-sideband measurement scheme described in Ref.~\cite{Clerk:2008}. If the optical cavity is driven by an amplitude-modulated signal with the appropriate phase, decomposing the cavity operator into the sum of classical and quantum parts $\hat{a}=\alpha+\delta\hat{a}$, moving into the frame rotating with the cavity and mechanical resonator, and making the rotating-wave approximation, the system Hamiltonian takes the form:
\begin{equation}\label{eq:H0}
    \hat{H}_0=-\hbar G\hat{X}_a\hat{Q}\; ,
\end{equation}
where $G$ is the renormalized optomechanical coupling strength, and we have introduced the cosine quadratures of the cavity $(\delta\hat{a}+\delta\hat{a}^\dagger)/\sqrt{2}=\hat{X}_a\cos(\omega_\mathrm{cav} t)+\hat{Y}_a\sin(\omega_\mathrm{cav} t)$ and the mechanical resonator $\hat{x}=\hat{Q}\cos(\omega_m t)+\hat{P}\sin(\omega_m t)$.

We assume that the system is coupled to an environment, modeled as a heat bath. The coupling to the environment introduces thermal damping, as well as quantum noise. Using standard input-output theory~\cite{Gardiner-Collett-input-output,Zoller-Gardiner-book}, we may derive the input-output equations for the cavity and mechanical resonator operators, as well as the cavity and mechanical resonator input (noise) fields~\cite{Aspelmeyer:2014}. The system operators coupling to the input noise fields are
\begin{equation}\label{eq:L}
    \hat{L}_\mathrm{cav}=\sqrt{\kappa}\delta\hat{a}\; , \qquad \hat{L}_m=\sqrt{\gamma}\hat{b}\; .
\end{equation}
Here, $\kappa$ and $\gamma$ are the cavity and mechanical damping rates, respectively. Hamiltonian~\eqref{eq:H0} and operators~\eqref{eq:L} define the SLH-coefficients~\cite{Gough:networks,Gough:series-product,Combes:2017} of the system (see Appendix~\ref{sec:quantum Kalman} for more details).

We use a (continuous-time) quantum Kalman filter to estimate system operators~\cite{Belavkin:1987,Belavkin:1989,Belavkin:1994}. The estimate of an operator may be used to apply measurement-based feedback. The Kalman filter provides the optimal least squares estimator for a variable conditional on past observations. The filtered estimate of an operator $\hat{X}$ based on the quantum measurement, denoted as $\pi_t(\hat{X})$, is a projection of the operator into the measurement output algebra. $\pi_t(\hat{X})$ commutes with all system operators, so it may be treated as a scalar.

The cavity output field $\hat{A}_\mathrm{out}$ is correlated with the state of the oscillator, so we may use it to perform a back-action evasion measurement of a mechanical quadrature~\footnote{Typically, feedback requires amplifying the signal from the quadrature measurement. We assume a phase-sensitive amplifier deep in the quantum limit which does not add any appreciable amount of noise.}. Specifically, the homodyne measurement process
\begin{equation}
    \hat{Y}=-i\hat{A}_\mathrm{out}+i\hat{A}_\mathrm{out}^\dagger
\end{equation}
provides information about the $\hat{Q}$ mechanical quadrature~\cite{Clerk:2008}. We apply a quantum Kalman filter over $\hat{Y}$ to provide estimates for system operators $\pi_t(\hat{X})$. Assuming that the input field of the cavity is in its vacuum state, the quantum Kalman filter satisfies the Belavkin-Kushner-Stratonovich (BKS) equation~\cite{Bouten:2007,bouten2008separation}.

The BKS equation for the estimate of $\hat{Q}$ quadrature is
\begin{equation}\label{eq:pi_t}
    d\pi_t(\hat{Q})=-\frac{\gamma}{2}\pi_t(\hat{Q})dt+\sqrt{2\kappa}\mathcal{V}_{Y_a,Q}dI\; ,
\end{equation}
where $\mathcal{V}_{Y_a,Q}=\pi_t(\hat{Y}_a\hat{Q})-\pi_t(\hat{Y}_a)\pi_t(\hat{Q})$. Equation~\eqref{eq:pi_t} updates the estimate of $\hat{Q}$ using system dynamics, described by the first term, and a correction driven by the innovations process $I(t)$, where $I(t)$ describes the difference between the present measurement of $\hat{Y}(t)$ and its prediction:
\begin{equation}\label{eq:I_t}
    dI=d\hat{Y}-\sqrt{2\kappa}\pi_t(\hat{Y}_a)dt\; .
\end{equation}
The equations for the other system operator estimates and their covariances are given in Eq.~\eqref{eq:pi}.

\subsection{Optical PID feedback squeezing}\label{sec:PID}
The high-sensitivity cavity readout measurement can be used to squeeze and control the mechanical degrees of freedom via active feedback. In Ref.~\cite{Clerk:2008}, it was shown that a proportional feedback force $F_\mathrm{fb}(t)=\alpha_P\gamma\pi_t(\hat{Q})\sin(\omega_m t)$ (in the laboratory frame) results in an additional effective damping of the $\hat{Q}$ quadrature, so that the total damping is $\gamma+\alpha_P\gamma/2$, where $\alpha_P$ is the proportional feedback strength. In this work, we generalize the feedback force in Ref.~\cite{Clerk:2008} to account for integral and derivative terms. For this purpose, we use a quantum Kalman filter based PID controller, developed in Ref.~\cite{Gough:2020}. We model the PID feedback via coefficients $\alpha_P$, $\alpha_I$ and $\alpha_D$. In addition, we also account for a setpoint force $F_\mathrm{sp}(t)=-\gamma r(t)\sin(\omega_m t)$, that may be used to drive the quadrature to the desired value $r(t)$ by applying the PID control over the error signal $e(t)=r(t)-\pi_t(\hat{Q})$.

Transforming the feedback force into the rotating frame and making the rotating-wave approximation, the PID feedback Hamiltonian $\hat{H}_\mathrm{PID}=\hat{H}_\mathrm{P}+\hat{H}_\mathrm{I}+\hat{H}_\mathrm{D}$ takes the form
\begin{subequations}\label{eq:HPID}
\begin{equation}
    \hat{H}_\mathrm{P}=\hbar\frac{\alpha_P\gamma}{2}(r(t)-\pi_t(\hat{Q}))\hat{P}
\end{equation}
\begin{equation}
    \hat{H}_\mathrm{I}=\hbar\frac{\alpha_I\gamma^2}{4}\int^t dt'(r(t')-\pi_{t'}(\hat{Q}))\hat{P}
\end{equation}
\begin{equation}\label{eq:HD}
    \hat{H}_\mathrm{D}=-\hbar\alpha_D\frac{d\pi_t(\hat{Q})}{dt}\hat{P}\; .
\end{equation}
\end{subequations}
We do not consider the derivative term over the setpoint in Eq.~\eqref{eq:HD}, as abrupt changes in the setpoint lead to large spikes in the derivative term. The derivative term of the feedback Hamiltonian~\eqref{eq:HD}, involves the formal derivative of $\pi_t(\hat{Q})$. This term lies outside the SLH-formalism, as it would introduce derivatives of stochastic terms in the filtered estimate equations. However, it can be transferred to the coupling operators~\eqref{eq:L}, as described in Appendix~\ref{sec:derivative}.

In practical realizations the performance of derivative feedback is typically limited by experimental imperfections~\cite{Li:2006}. Finite measurement bandwidth and feedback delays reduce the accuracy of derivative estimation and limit the controller's ability to respond to rapid changes in the measured signal. Furthermore, differentiation amplifies high-frequency fluctuations, making the feedback particularly sensitive to measurement and electronic noise. A common approach to mitigate this issue is to apply a low-pass filter to the measured signal before differentiation, which slows the response and distorts the derivative as a trade-off. Consequently, the filter bandwidth must be chosen carefully to balance noise suppression against control performance. Our ideal model works when the bandwidth exceeds the inverse of the fastest time scale of the dynamics in the rotating frame.

The feedback Hamiltonian~\eqref{eq:HPID} modifies the BKS equation for the $\hat{Q}$ quadrature to the form:
\begin{multline}\label{eq:Q}
    d\pi_t(\hat{Q})=\frac{1}{1+\alpha_D}\Bigg[\Big(-\frac{\gamma}{2}\pi_t(\hat{Q})+\frac{\alpha_P\gamma}{2}(r(t)-\pi_t(\hat{Q}))\\
    +\frac{\alpha_I\gamma^2}{4}\int^t dt'(r(t')-\pi_{t'}(\hat{Q}))\Big)dt+\sqrt{2\kappa}\mathcal{V}_{Y_a,Q}dI\Bigg]\; .
\end{multline}
Applying a known force to the system cannot change the observer's estimate uncertainty, so the feedback will not modify the estimate covariances.

\section{Squeezing of a mechanical degree of freedom}
As mentioned above, proportional feedback leads to the squeezing of the $\hat{Q}$ quadrature. In the following, we study how it can be combined with integral and derivative feedback to enhance the squeezing. The estimate variance $\mathcal{V}_{Q}=\pi_t(\hat{Q}^2)-\pi_t(\hat{Q})^2$ provides the conditional variance of the observable, i.e., the variance in a particular run of the experiment. True squeezing is achieved when the unconditional variance $V_Q$ drops below the zero-point value. The term unconditional means that the variance is not conditioned on the measurement record~\cite{Clerk:2022,Clerk:2008}. The unconditional variance is related to $\mathcal{V}_{Q}$ via
\begin{equation}\label{eq:VX}
    V_Q =\langle \hat{Q}^2\rangle-\langle\hat{Q}\rangle^2=\mathcal{V}_Q+\langle \pi_t(\hat{Q})^2\rangle-\langle \pi_t(\hat{Q})\rangle^2\; .
\end{equation}
The $\langle \pi_t(\hat{Q})^2\rangle-\langle \pi_t(\hat{Q})\rangle^2$ term in Eq.~\eqref{eq:VX} describes the excess noise originating from the averaging of the input field fluctuations.

All expectation values obey deterministic linear ordinary differential equations. We first solve the estimate covariances, then the linear and quadratic estimate averages, and finally obtain the unconditional variances numerically. All relevant equations are given in Appendix~\ref{sec:averages}. The code used to solve the equations is available at~\cite{programs}. We also derive the conditional and the total (unconditional) variances analytically in the stationary regime to illustrate the effect of the PID action. Due to the inherent damping of the optomechanical system, the system will evolve into a stationary regime on a timescale of the order of a few $\gamma^{-1}$.

\begin{figure}[t!]
    \centering
    \includegraphics[width=0.99\columnwidth]{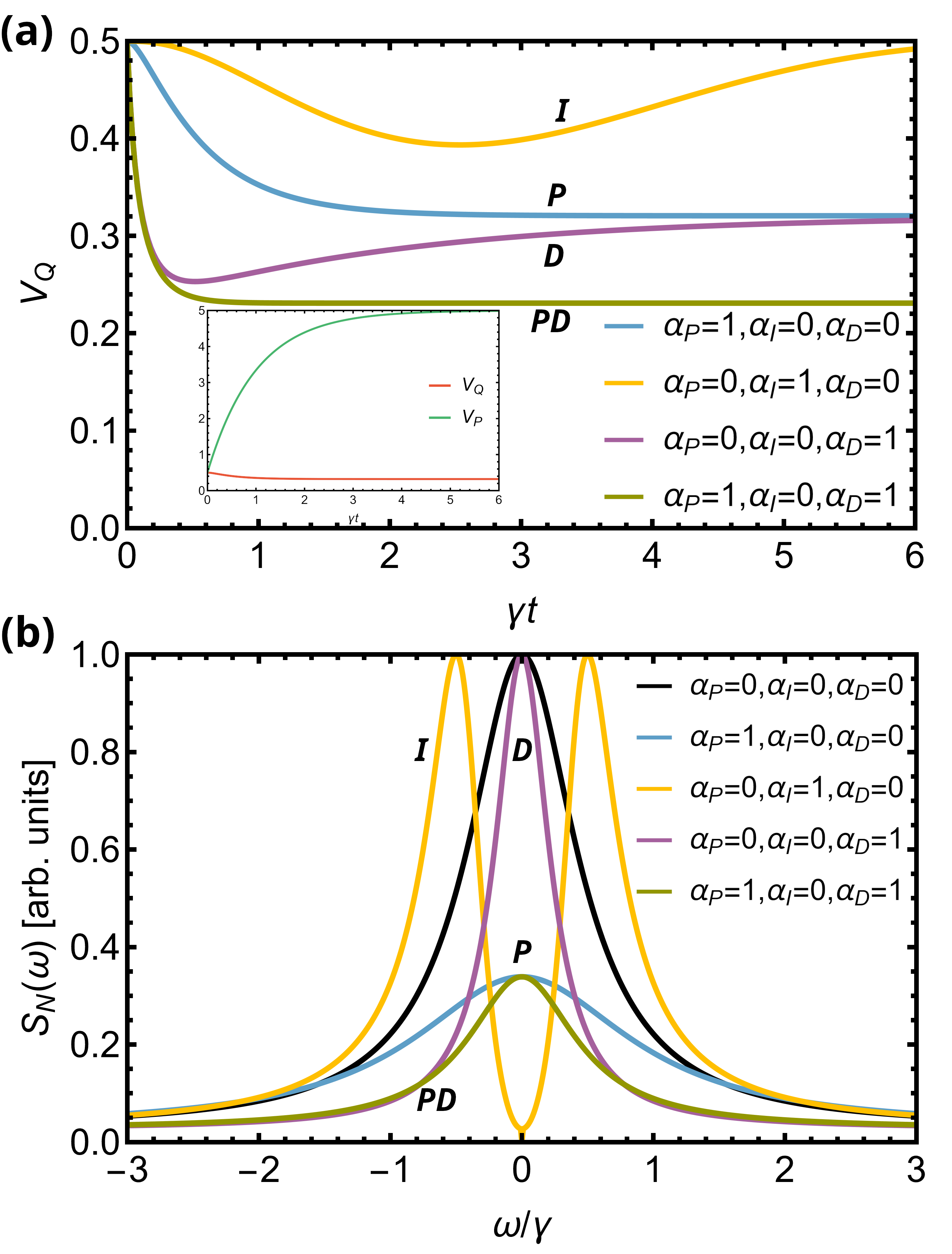}
    \caption{(a) Effect of PID feedback on the unconditional mechanical quadrature variances. The inset shows the dynamics of the variances of $\hat{Q}$ and $\hat{P}$ for proportional action $\alpha_P=1$. The main panel shows the squeezing of $\hat{Q}$ for different PID parameters. (b) Noise spectral density $S_N(\omega)$ of the output signal of the measurement for different PID parameters.}
    \label{fig:variance}
\end{figure}

Typically, the mechanical damping and the optomechanical coupling are much weaker than the cavity damping $G,\gamma \ll \kappa$. In this regime, we may expand the equations in Appendix~\ref{sec:averages} to find an analytical solution for the conditional and unconditional uncertainties in the steady state. Expanding the equations to first order in $\gamma/\kappa$ and $G/\kappa$, the filtered estimate variances of the $\hat{Q}$ and $\hat{P}$ operators, which provide the conditional variance, are
\begin{subequations}\label{eq:V}
\begin{equation}
    \mathcal{V}_Q^{ss}=n_\mathrm{th}+\frac{1}{2}-4n_\mathrm{BA}\left(n_\mathrm{th}+\frac{1}{2}\right)^2
\end{equation}
\begin{equation}
    \mathcal{V}_P^{ss}=n_\mathrm{th}+\frac{1}{2}+n_\mathrm{BA}\; ,
\end{equation}
\end{subequations}
where $n_\mathrm{th}=(e^{\hbar\omega_m/k_B T}-1)^{-1}$ describes the effect of thermal noise on the variance, $n_\mathrm{BA}=\frac{2G^2}{\gamma\kappa}$ is the measurement back-action, and the $ss$ superscript denotes the steady state.

The unconditional variances are obtained by averaging the random motion of the quadratures over multiple realizations of the experiment. The stochastic nature of the process appears as excess noise, as given by Eq.~\eqref{eq:VX}:
\begin{subequations}\label{eq:Var}
\begin{multline}
    V_Q^{ss}=n_\mathrm{th}+\frac{1}{2}-4n_\mathrm{BA}\left(n_\mathrm{th}+\frac{1}{2}\right)^2\\
    \cdot\left(1-\frac{1}{(1+\alpha_P)(1+\alpha_D)}\right)
\end{multline}
\begin{equation}
    V_P^{ss}=n_\mathrm{th}+\frac{1}{2}+n_\mathrm{BA}\; .
\end{equation}
\end{subequations}

In Fig.~\ref{fig:variance}(a), we show the unconditional variances of the mechanical quadratures, $\hat{Q}$ and $\hat{P}$. Here and below, we use the parameters $\kappa=0.1\omega_m$, $\gamma=10^{-5}\omega_m$ and $G=1.5 \cdot 10^{-3}\omega_m$, which are within the range of typical optomechanical systems. We assume that the system is initially in its ground state ($n_\mathrm{th}=0$), where $V_Q=V_P=1/2$. At $t=0$, we start the continuous measurement of the system and apply the measurement-based feedback. As shown in the inset of Fig.~\ref{fig:variance}(a), the measurement of $\hat{Q}$ induces the back-action on the $\hat{P}$ quadrature. As shown in Eqs.~(\ref{eq:V}-\ref{eq:Var}), in the good cavity limit $\kappa\ll\omega_m$, the measurement back-action is quantified by $n_\mathrm{BA}=\frac{2G^2}{\gamma\kappa}$~\cite{Clerk:2008}. For the used parameters, $n_\mathrm{BA}=4.5$, leading to a variance of $V_P=5$ in the stationary regime. The information gained from the measurement, and the subsequent back-action of $\hat{P}$, enables the squeezing of $\hat{Q}$.

In the main panel of Fig.~\ref{fig:variance}(a) we compare how the different feedback terms affect the squeezing of $\hat{Q}$. The \textit{proportional feedback} (blue line) induces an effective damping rate of $\alpha_P\gamma/2$ [see Eq.~\eqref{eq:Q}]. As a result, it suppresses the total unconditional variance, where the squeezing increases monotonously with $\alpha_P$~\cite{Clerk:2008}. The \textit{derivative feedback} also diminishes the excess noise, leading to a net unconditional squeezing (purple line), as shown by Eq.~\eqref{eq:Var}. The \textit{integral feedback} (yellow line) induces a transient squeezing, but vanishes in the stationary regime. The integral action is proportional to $\sigma=\int^t dt'\langle\pi_t(\hat{Q})\pi_{t'}(\hat{Q})\rangle$. In the absence of a setpoint, $\sigma$ vanishes in the stationary regime, as shown by Eq.~\eqref{eq:sigma}.

In conclusion, proportional and derivative feedback lead to a net squeezing of $\hat{Q}$, with the unconditional variance decreasing monotonously with $\alpha_P$ and $\alpha_D$. As shown in Fig.~\ref{fig:variance}(a), we may combine the PD actions to enhance the squeezing of $\hat{Q}$. Moreover, the squeezing occurs at a higher speed than when using either the P or D actions.

Next, we study how the PID terms suppress the noise in the output signal of the measurement. The homodyne current is related to the output process $\hat{Y}$ as
\begin{equation}
    \hat{J}=\frac{d\hat{Y}}{dt}\; .
\end{equation}
The noise spectral density, describing the noise distribution across the frequency spectrum, is given by
\begin{equation}\label{eq:SN}
    S_N(\omega)=\int \frac{d\omega'}{2\pi}\langle \hat{J}(\omega)\hat{J}(\omega')\rangle
\end{equation}
where $J(\omega)$ is the Fourier transform of the homodyne current. We consider the equivalent classical formulation by replacing the mean values of the operators and their estimates with classical dynamical variables, and the noise terms with classical white noise fields~\cite{Doherty:1999}. We solve the coupled Heisenberg-Langevin equations for the operators and BKS equations for the estimates, given in Appendix~\ref{sec:averages}, in terms of the noise fields. Finally, we evaluate the correlators in Eq.~\eqref{eq:SN} to obtain the noise spectral density.

In Fig.~\ref{fig:variance}(b), we compare the suppression of noise in the output signal for different PID parameters. Without feedback (black line), the output spectrum consists of a constant contribution from the measurement noise and a Lorentzian contribution arising from the system dynamics. Proportional feedback (blue line) suppresses the noise across all frequencies, potentially enhancing the sensitivity of the measurement. As described in Ref.~\cite{Clerk:2008}, the area under the Lorentzian contribution is related to the unconditional variance. 

As discussed above, integral feedback alone does not lead to stationary squeezing. This is reflected in the corresponding noise spectral density (yellow line): the total noise power is conserved but redistributed across frequencies. This redistribution may be useful for measurements targeting specific frequency ranges. Derivative feedback (purple line), in contrast, leaves the zero-frequency noise unchanged while narrowing the noise bandwidth. Consequently, the combined use of PD feedback results in a further reduction in noise (green line).

\section{Control of a quadrature}

\begin{figure}[t!]
    \centering
    \includegraphics[width=0.99\columnwidth]{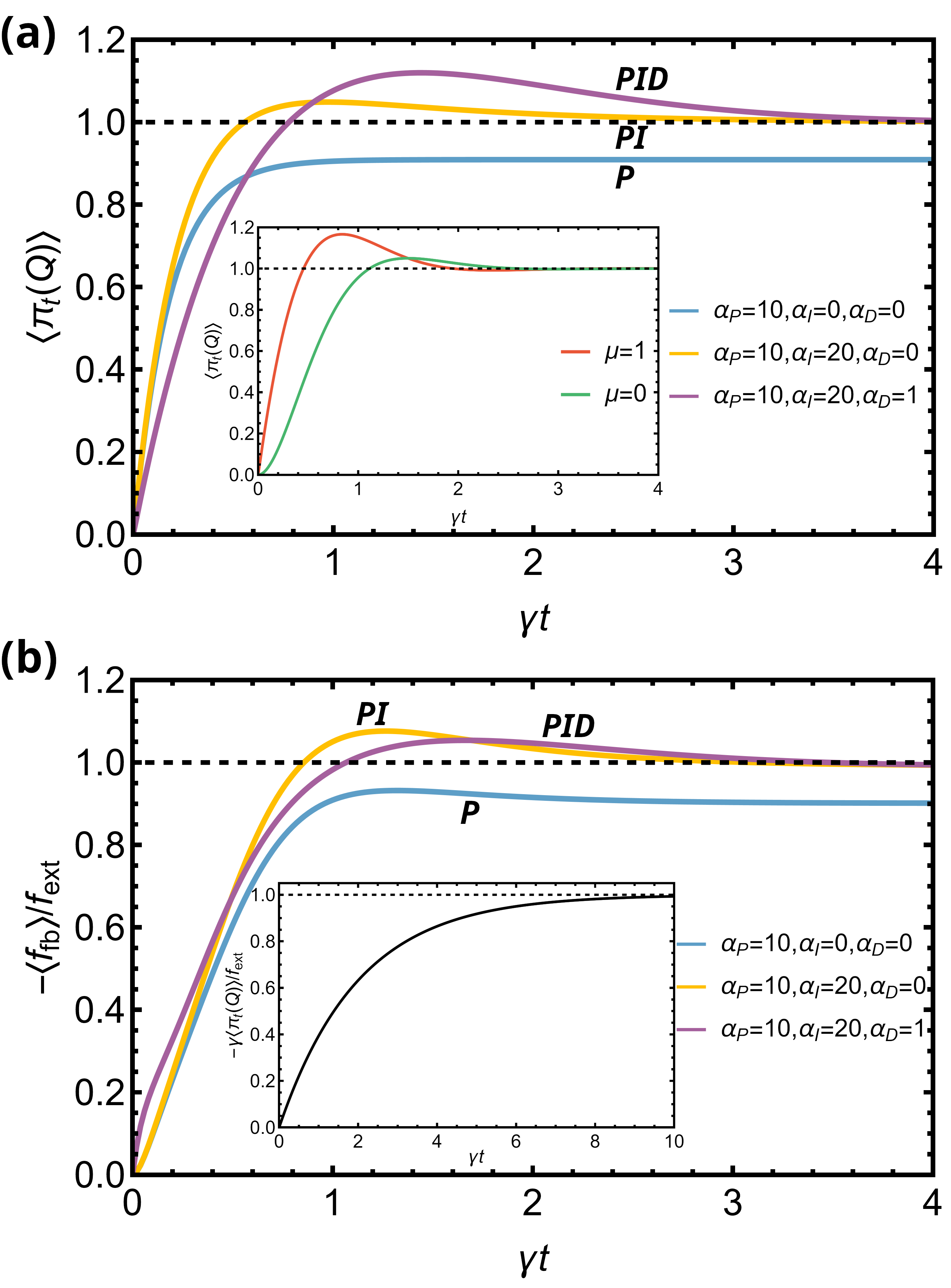}
    \caption{(a) Control of the $\hat{Q}$ quadrature for different PID parameters. The dashed line corresponds to the setpoint $r(t)=\theta(t)$. The inset shows the evolution of the quadrature for $\alpha_P=7$, $\alpha_I=33.6$ and $\alpha_D=0$, with ($\mu=1$) and without ($\mu=0$) proportional control over the setpoint. (b) Tracking of an external force $f_\mathrm{ext}$ (dashed line) switched on at $t=0$ for different PID parameters. The inset shows the tracking of the external force through quadrature measurement in a system without feedback. Note the different horizontal scales in the inset and in the main figure.}
    \label{fig:control_quadrature}
\end{figure}

In linear systems, the dynamics of the system is determined by the zeroes and poles of its transfer function $G(s)=Y(s)/R(s)$, where $R(s)$ and $Y(s)$ are the input and output signals in the Laplace domain. In the setting of Fig.~\ref{fig:sketch}, the output signal is $\pi_t(\hat{Q})$. Taking the Laplace transform of Eq.~\eqref{eq:Q}, for homogeneous initial conditions, we arrive at
\begin{equation}\label{eq:G}
    G(s)=\frac{s\frac{\alpha_P\gamma}{2}+\frac{\alpha_I\gamma^2}{4}}{s^2(1+\alpha_D)+s\frac{(1+\alpha_P)\gamma}{2}+\frac{\alpha_I\gamma^2}{4}}\; .
\end{equation}
$\langle\pi_t(\hat{Q})\rangle$ satisfies a second-order differential equation~\eqref{eq:Q}, so the transfer function contains two poles. The position of the zeros and poles in the transfer function determine the dynamics of the system, so the PID parameters may be tuned to transform the denominator into some desired second-order system of the form $s^2+2\varsigma\omega_ns+\omega_n^2$.

We start by studying how to induce a displacement of a quadrature to a given setpoint $r(t)$. In Fig.~\ref{fig:control_quadrature}(a) we show the step response of the system ($r(t)=\theta(t)$) for illustrative values of the PID parameters. Using proportional control alone (blue line) results in a steady-state error between the setpoint and the output $e(t)=r(t)-\pi_t(\hat{Q})$. This steady-state error arises due to the finite mechanical loss. Using the final value theorem~\cite{Laplace-transform-book} for transfer function~\eqref{eq:G}, the output for a step function is $\pi_\infty(\hat{Q})=\alpha_P/(1+\alpha_P)$. As usual in second order systems, increasing the proportional feedback increases the speed of the response and reduces the steady-state offset. The integral term applies a control term that depends on the cumulative value of $e(t)$. Adding integral control (yellow line) eliminates the steady-state error, as the integral control only vanishes once the error is eliminated. However, integral feedback introduces some overshoot, slowing down the response. $e(t)$ is very large at the beginning, so the integral action induces a large response, leading to the overshoot. Adding the derivative action (purple line) does not provide any benefit to quadrature control, as the system becomes slower and more unstable. This is in contrast to the usual role of the derivative action, which is generally used to reduce overshoot and increase stability margin. Therefore, a PI controller is sufficient for this application.

PID parameters may also be tuned to satisfy speed and stability specifications, for example, to initialize the quadrature in a force-sensing experiment within the desired time frame. In Appendix~\ref{sec:tuning}, we describe the tuning procedure used to obtain the time response shown in the inset of Fig.~\ref{fig:control_quadrature}(a). Here, we consider an additional tuning parameter $\mu$, which sets the weight of the setpoint in the proportional feedback.

\section{Measurement of a weak force}\label{sec:measuring}

A sinusoidal force $F_\mathrm{ext}(t)=f_\mathrm{ext}(t)\sin{(\omega_m t)}$, where $f_\mathrm{ext}(t)$ is slowly time dependent, with a time constant much longer than $1/\omega_m$, induces a steady-state displacement of the quadrature $\hat{Q}$. If the displacement is greater than the measured quadrature uncertainty, the measurement may be used to detect the force, as explained in Appendix~\ref{sec:force}. The mechanical resonator reacts to sudden changes of the amplitude of the force in an exponential trend, with a time constant $2/\gamma$, as shown in the inset of Fig.~\ref{fig:control_quadrature}(b). In the following, we study how PID feedback may be used to decrease the system's response time for sensing applications. Unlike the setpoint signal $r(t)$, we generally do not have access to the external force, so we may not directly include it in the Kalman filter model. Instead, we augment the Kalman filter to include the estimate of the classical variable $f_\mathrm{ext}(t)$, as described in Appendix~\ref{sec:f_ext}. Moreover, we may not directly apply the PID control to the external force. This means that the PID feedback cannot be used to drive the $\hat{Q}$ quadrature faster to its target value from which we would obtain the value of $f_\mathrm{ext}$.

One possibility is to use the estimate of the force provided by the Kalman filter. Yet, the speed at which the estimate of the force gets updated is determined by the innovations process of the filter. This can lead to inaccurate estimates of the force in the transient regime. Instead, we use the feedback force, given by the output signal of the PID block [see Fig.~\ref{fig:sketch}(b)]
\begin{equation}
    f_\mathrm{fb}(t)=2\left(\frac{\alpha_P\gamma}{2}\pi_t(\hat{Q})+\frac{\alpha_I\gamma^2}{4}\int^t dt'\pi_{t'}(\hat{Q})+\alpha_D\frac{d\pi_t(\hat{Q})}{dt}\right)\; ,
\end{equation}
as a measure of the external force. In the steady state, the sum of the external, feedback and damping forces is zero, so $f_\mathrm{fb}\approx-f_\mathrm{ext}$. The PID control may be tuned to modify the dynamics of $f_\mathrm{fb}$, providing a faster and more accurate sensing of the external force.

In Fig.~\ref{fig:control_quadrature}(b), we consider an external force that starts acting on the system at $t=0$. We plot the feedback force $f_\mathrm{fb}$ for illustrative values of the PID parameters. Similar to the quadrature control, in the case of proportional feedback (blue line) there is a steady-state error in the estimation of the external force. This can be solved by adding integral feedback (yellow line). In the absence of a setpoint $r=0$, the integral feedback drives the system to a state with $\langle\hat{Q}\rangle=0$ where the damping vanishes, so the relation $f_\mathrm{ext}=-f_\mathrm{fb}$ becomes exact. Regarding the derivative term (purple line), it reduces the overshoot of the response, barely increasing the settling time. This thus shows that using the feedback force as a measure of $f_\mathrm{ext}$ allows for a faster way to track an external force, compared to the usual quadrature measurement. In other words, it can thus adapt to changes in the force that are smaller than the intrinsic speed limit set by the damping constant.

\section{Discussion}\label{sec:conclusion}
We have studied the squeezing and control of a mechanical quadrature in an optomechanical system via PID feedback. We have provided a fully quantum treatment of the non-demolition measurement and feedback scheme, accounting for dynamical and observational noise, and deriving analytical equations for the system operator estimates and variances. We show how the integral and derivative feedback may be combined with the usual proportional feedback to enhance the transient and stationary response. In particular, we show that the derivative action may be used to speed up and enhance the squeezing of a quadrature. We demonstrate that measurement-based feedback can be employed to drive the quadrature to a desired setpoint. A combined use of the proportional and integral actions offers a way to eliminate the steady-state offset. Moreover, frequency domain analysis offers a way to tune the PID parameters for precise control of the quadrature. Finally, PID control allows for a faster readout in a force-sensing experiment.

The measurement of a mechanical quadrature can be used for an ultra-sensitive force detection~\cite{Caves:1980,Clerk:2010,Gavartin:2012,Arvidsson:2024}, for instance, for gravitational wave detection or testing quantum gravity by measuring the gravitational force from a source mass that is in a spatially separated superposition. In this case, the measurement precision is fundamentally limited by quadrature fluctuations. Measurement-based feedback can be applied to reduce these fluctuations and squeeze the quadrature below the zero-point level. The PID feedback also offers a way to enhance the speed and magnitude of the squeezing, allowing for faster and more accurate measurements. Future work could explore the application of PID control in more general open quantum systems, including qubit-based systems.

\begin{acknowledgments}
We thank M. Sillanpää for fruitful discussions. This project has received funding from the European Research Executive Agency under the European Union’s Horizon TMA MSCA Postdoctoral Fellowships – European Fellowships action Grant agreement No 101202316. The work is also part of the Finnish Centre of Excellence in Quantum Materials (QMAT).
\end{acknowledgments}

\section*{Data availability}
The data that support the findings of this study were generated by numerical simulations. The source code and the generated data are publicly available at~\cite{programs}.

\appendix
\numberwithin{equation}{section}
\renewcommand{\thesubsection}{\arabic{subsection}}
\section{Quantum Kalman filter}\label{sec:quantum Kalman}
In this section, we introduce the SLH formalism~\cite{Gough:networks,Gough:series-product,Combes:2017} used to study the optomechanical resonator under measurement-based feedback. The SLH formalism is a synthesis of the quantum stochastic calculus developed
by Hudson-Parthasarathy in 1984~\cite{Hudson:1984,Parthasarathy-book} and the quantum input-output theory developed by Gardiner and Collett~\cite{Gardiner-Collett-input-output,Zoller-Gardiner-book}.

We consider an open quantum system driven by thermal bosonic input fields $\hat{b}_{\mathrm{in},i}$, with $d\hat{B}_i=\hat{b}_{\mathrm{in},i}dt$. We consider uncorrelated white-noise input ensembles, where the non-zero Itô products are: $d\hat{B}^\dagger_i d\hat{B}_j=n_i\delta_{ij}dt$ and $d\hat{B}_i d\hat{B}^\dagger_j=(n_i+1)\delta_{ij}dt$~\cite{Hudson:1985}. The general form of the quantum stochastic differential equation for this adapted unitary process $\hat{U}$ is~\cite{Combes:2017,Gough:2025}
\begin{multline}\label{eq:U}
    d\hat{U}(t)=\Bigg\{-\left(\sum_i\frac{n_i+1}{2}\hat{L}_i^\dagger\hat{L}_i+\frac{n_i}{2}\hat{L}_i\hat{L}_i^\dagger+i\hat{H}\right)dt\\
    +\sum_i\hat{L}_id\hat{B}_i^\dagger(t)-\sum_i\hat{L}^\dagger_id\hat{B}_i(t)\Bigg\}\hat{U}(t)
\end{multline}
where $\hat{H}$ is self-adjoint, and $\hat{B}_i$ and $\hat{B}^\dagger_i$ are annihilation and creation processes, respectively. $\hat{L}_i$ and $\hat{H}$ (together with the operators $\hat{S}_{ij}$ in systems subject to gauge processes), are termed the Hudson-Parthasarathy parameters, also known as the SLH-coefficients~\cite{Gough:networks,Gough:series-product,Combes:2017}.

The time evolution of a system operator $\hat{X}$ is given by the unitary transformation:
\begin{equation}
    \hat{X}(t)=\hat{U}^\dagger\hat{X}(0)\hat{U}
\end{equation}
Using Eq.~\eqref{eq:U} and applying the rules of Itô calculus, we obtain the Heisenberg-Langevin equation:
\begin{equation}\label{eq:j}
    d\hat{X}=\mathcal{L}\hat{X}dt+\sum_i \mathcal{L}_{i0}\hat{X}d\hat{B}^\dagger_i(t)+\sum_i \mathcal{L}_{0i}\hat{X}d\hat{B}_i(t)
\end{equation}
with
\begin{subequations}\label{eq:Lindbladian}
\begin{multline}
    \mathcal{L}\hat{X}=i[\hat{H},\hat{X}]+\sum_i \bigg[\frac{n_i+1}{2}\left(\hat{L}_i^\dagger[\hat{X},\hat{L}_i]+[\hat{L}_i^\dagger,\hat{X}]\hat{L}_i\right)\\
    +\frac{n_i}{2}\left(\hat{L}_i[\hat{X},\hat{L}_i^\dagger]+[\hat{L}_i,\hat{X}]\hat{L}_i^\dagger\right)\bigg]
\end{multline}
\begin{equation}
    \mathcal{L}_{i0}\hat{X}=[\hat{X},\hat{L}_i]
\end{equation}
\begin{equation}
    \mathcal{L}_{0i}\hat{X}=[\hat{L}^\dagger_i,\hat{X}]
\end{equation}
\end{subequations}
Comparing Eqs.~(\ref{eq:j}-\ref{eq:Lindbladian}) to the input-output formalism~\cite{Gardiner-Collett-input-output}, we may identify $\hat{H}$ as the system Hamiltonian, $\hat{L}$ as the system operators coupling to the input noise fields and $\mathcal{L}$ as the Lindbladian.

Similarly, the output fields are given by $\hat{B}_{\mathrm{out},i}=\hat{U}^\dagger\hat{B}_i\hat{U}$. Application of Itô calculus yields
\begin{equation}\label{eq:dB_out}
    d\hat{B}_{\mathrm{out},i}=d\hat{B}_i(t)+\hat{L}_i dt\; .
\end{equation}

The output fields may be used to perform a non-demolition measurement. For instance, in a homodyne measurement, we may measure the quadrature process
\begin{equation}\label{eq:Y}
    \hat{Y}(t)=e^{i\theta}\hat{B}_{\mathrm{out},i}+e^{-i\theta}\hat{B}^\dagger_{\mathrm{out},i}\; .
\end{equation}
$\hat{Y}(t)$ is a quantum Wiener process~\cite{Lim:2024} satisfying $d\hat{Y}^2=(2n_i+1)dt$. It is a self-commuting process, forming a commutative von Neumann algebra $\mathcal{N}_t=\mathrm{vN}\{\hat{Y}(s):0 \leq s \leq t\}$. Moreover, the Heisenberg picture system operators $\hat{X}$ commute with past values of the output field $[\hat{X}(t),\hat{Y}(s)]$, for $s \leq t$, so it is compatible with the measurement algebra $\mathcal{N}_t$. Therefore, we may use measurements up to current time to estimate (filter) any system operator $\hat{X}$~\cite{Belavkin:1983,Belavkin:1994}.

Here we use a (continuous-time) quantum Kalman filter to estimate system operators~\cite{Belavkin:1987,Belavkin:1989,Belavkin:1994}. The Kalman filter provides the optimal least squares estimator for a variable conditional on past observations. The filtered estimate of $\hat{X}$ based on the quantum measurement, denoted as $\pi_t(\hat{X})$, is a projection of the operator into the measurement output algebra $\mathcal{N}_t$. Since all elements in $\mathcal{N}_t$ commute, $\pi_t(\hat{X})$ may be treated as a scalar, and the noise terms in the estimation may be treated as white noise.

The least squares property of the Kalman filter implies that the estimate of $(\pi_t(\hat{X})-\hat{X}(t))y(t)$ vanishes for all $y\in\mathcal{N}_t$. Taking
the Itô differential of this identity for the particular case $dy=f(t)y(t)d\hat{Y}$, with $f(t)$ an arbitrary integrable function, we obtain the equation satisfied by the filter. For a measured field given by~\eqref{eq:Y}, assuming that the input field is in its vacuum state, the quantum Kalman filter satisfies the Belavkin-Kushner-Stratonovich (BKS) equation~\cite{Bouten:2007,bouten2008separation}:
\begin{equation}\label{eq:BKS}
    d\pi_t(\hat{X})=\pi_t(\mathcal{L}\hat{X})dt+\mathcal{G}(\hat{X})dI(t); ,
\end{equation}
with
\begin{equation}
    \mathcal{G}(\hat{X})=\pi_t(e^{i\theta}\hat{X}\hat{L}+e^{-i\theta}\hat{L}^\dagger\hat{X})-\pi_t(\hat{X})\pi(e^{i\theta}\hat{L}+e^{-i\theta}\hat{L}^\dagger)
\end{equation}
and $I(t)$ is the innovations process, which describes the difference between what we observe $\hat{Y}$ and our prediction based on past measurements:
\begin{equation}\label{eq:I}
    dI(t)=d\hat{Y}-\pi_t(e^{i\theta}\hat{L}+e^{-i\theta}\hat{L}^\dagger)dt\; .
\end{equation}
$I(t)$ has the statistics of a Wiener process and thus $dI^2=dt$.

\subsection{Filtering of an optomechanical system}
We assume that the heat bath couples resonantly with the cavity and the resonator, such that the nonzero products of the Itô increments are
\begin{subequations}
\begin{equation}
    d\hat{A}d\hat{A}^\dagger=dt
\end{equation}
\begin{equation}
    d\hat{B}^\dagger d\hat{B}=n_\mathrm{th}dt
\end{equation}
\begin{equation}
    d\hat{B}d\hat{B}^\dagger=(n_\mathrm{th}+1)dt
\end{equation}
\end{subequations}
where $\hat{A}$ and $\hat{B}$ are the cavity and resonator input fields, respectively, and $n_\mathrm{th}=(e^{\hbar\omega_m/k_B T}-1)^{-1}$ describes the thermal state of the bath field coupling to the mechanical resonator. Here, we assume that the cavity resonance frequency is much higher than the temperature, $\hbar\omega_\mathrm{cav} \gg k_B T$.

The cavity output field is correlated with the state of the oscillator. Specifically, the measurement process
\begin{equation}
    \hat{Y}=-i\hat{A}_\mathrm{out}+i\hat{A}_\mathrm{out}^\dagger
\end{equation}
provides information about the $\hat{Q}$ mechanical quadrature~\cite{Clerk:2008}. From \eqref{eq:dB_out}, we obtain the relation between the output process and the input fields:
\begin{equation}
    d\hat{Y}=d\hat{Z}+\sqrt{2\kappa}\hat{Y}_a dt\; ,
\end{equation}
where $\hat{Z}=-i\hat{A}+i\hat{A}^\dagger$ is the input process. Here, we have introduced the quadratures of the cavity $(\delta\hat{a}+\delta\hat{a}^\dagger)/\sqrt{2}=\hat{X}_a\cos(\omega_\mathrm{cav} t)+\hat{Y}_a\sin(\omega_\mathrm{cav} t)$ and the mechanical resonator $\hat{x}=\hat{Q}\cos(\omega_m t)+\hat{P}\sin(\omega_m t)$.

Since Hamiltonian~\eqref{eq:H0} in the main text is quadratic in the system operators, Heisenberg dynamics are linear for the optical and mechanical quadratures. If the system is initially in a Gaussian state, we have a linear Gaussian filtering problem. This allows us to write the filtered estimate equations in terms of their covariances (see Appendix~\ref{sec:covariances}), which are deterministic functions. The BKS equations~\eqref{eq:BKS} for the system operators are:
\begin{subequations}\label{eq:pi}
\begin{equation}\label{eq:piQ}
    d\pi_t(\hat{Q})=-\frac{\gamma}{2}\pi_t(\hat{Q})dt+\sqrt{2\kappa}\mathcal{V}_{Y_a,Q}dI
\end{equation}
\begin{equation}
    d\pi_t(\hat{P})=\left(G\pi_t(\hat{X}_a)-\frac{\gamma}{2}\pi_t(\hat{P})\right)dt+\sqrt{2\kappa}\mathcal{V}_{Y_a,P}dI
\end{equation}
\begin{equation}
    d\pi_t(\hat{X}_a)=-\frac{\kappa}{2}\pi_t(\hat{X}_a)dt+\sqrt{2\kappa}\mathcal{V}_{X_a,Y_a}dI
\end{equation}
\begin{equation}
    d\pi_t(\hat{Y}_a)=\left(G\pi_t(\hat{Q})-\frac{\kappa}{2}\pi_t(\hat{Y}_a)\right)dt+\sqrt{2\kappa}\left(\mathcal{V}_{Y_a}-\frac{1}{2}\right)dI
\end{equation}
\end{subequations}
where $\mathcal{V}_{\hat{X},\hat{Y}}=\pi_t(\{\hat{X},\hat{Y}\}/2)-\pi_t(\hat{X})\pi_t(\hat{Y})$ are the filtered estimate covariances, and we omit the second subscript for variances $\mathcal{V}_X\equiv\mathcal{V}_{X,X}$. The innovations process $dI$ is obtained from Eq.~\eqref{eq:I}
\begin{equation}\label{eq:dI}
    dI=d\hat{Y}-\sqrt{2\kappa}\pi_t(\hat{Y}_a)dt=d\hat{Z}+\sqrt{2\kappa}(\hat{Y}_a-\pi_t(\hat{Y}_a))dt\; .
\end{equation}
Equations \eqref{eq:pi} and \eqref{eq:dI} correspond to Eqs.~\eqref{eq:pi_t} and~\eqref{eq:I_t} in the main text, respectively. They provide the evolution of the estimate of the quadratures in the absence of feedback. In Appendix~\ref{sec:averages}, we generalize the equations to account for measurement-based feedback.

\section{Treatment of the derivative feedback Hamiltonian}\label{sec:derivative}
The derivative term of the feedback Hamiltonian~\eqref{eq:HD} in the main text, involves the formal derivative of $\pi_t(\hat{Q})$. This term lies outside the class of model allowed by the SLH-formalism, as it would introduce derivatives of stochastic terms in the Hamiltonian. This term can be treated by transferring the fluctuating parts to the $L$ coefficient as~\cite{Wiseman:1994,Gough:2009}
\begin{equation}\label{eq:L_cav}
    \hat{L}_\mathrm{cav}=\sqrt{\kappa}\delta\hat{a}+\hat{F}_D\; ,
\end{equation}
and adding a correction to the Hamiltonian:
\begin{multline}\label{eq:HD0}
\hat{H}_{D}=\hbar\alpha_D\left(-\pi_t(\mathcal{L}_0\hat{Q})+\frac{2\kappa}{1+\alpha_D}\mathcal{V}_{Y_a,Q}\pi_t(\hat{Y}_a)\right)\hat{P}\\
+\frac{\hbar}{2i}(\hat{F}_D\hat{L}_{\mathrm{cav},0}-\hat{L}_{\mathrm{cav},0}^\dagger\hat{F}_D)\; ,
\end{multline}
where $\hat{F}_D$ is the self-adjoint process
\begin{equation}\label{eq:FD}
    \hat{F}_D=-\frac{\alpha_D}{1+\alpha_D}\sqrt{2\kappa}\mathcal{V}_{Y_a,Q}\hat{P}\; .
\end{equation}

The first term in Eq.~\eqref{eq:HD0} is obtained by replacing $d\pi_t(\hat{Q})/dt$ with its deterministic part [see Eq.~\eqref{eq:BKS}], with $\mathcal{L}_0$ the Lindbladian in the absence of derivative action. The second term is a correction resulting from the shift of the coupling operator~\cite{Gough:2009,Gough:2017}. Combining Eqs.~(\ref{eq:HD0}-\ref{eq:FD}), the Hamiltonian term describing the derivative action takes the form:
\begin{multline}\label{eq:HD2}
    \hat{H}_{D}=\hbar\frac{\alpha_D}{1+\alpha_D}\bigg(\frac{\gamma}{2}\pi_t(\hat{Q})-\frac{\alpha_P\gamma}{2}(r(t)-\pi_t(\hat{Q}))\\
    -\frac{\alpha_I\gamma^2}{4}\int^t dt'(r(t')-\pi_{t'}(\hat{Q}))+\kappa(2\pi_t(\hat{Y}_a)-\hat{Y}_a)\mathcal{V}_{Y_a,Q}\bigg)\hat{P}
\end{multline}
In Appendix~\ref{sec:averages}, for deriving the BKS equations for the system operators, we use the coefficient $L$ given by Eq.~\eqref{eq:L_cav} and the derivative Hamiltonian~\eqref{eq:HD0} instead of Hamiltonian~\eqref{eq:HD} introduced in the main text.

\section{Equations of motion under feedback}\label{sec:averages}
In this section, we provide the equations of motion for the quadratures and their variances under feedback. In the main text, we solve the coupled system of equations to compute the unconditional variance of the mechanical quadratures.

\subsection{Equations for the real quadratures}
The Heisenberg-Langevin equations~\eqref{eq:j} for the real quadratures are:
\begin{subequations}\label{eq:linear_real}
\begin{multline}
    d\hat{Q}=\Big(-\frac{\gamma}{2}\hat{Q}+\frac{1}{1+\alpha_D}\Big[\alpha_D\frac{\gamma}{2}\pi_t(\hat{Q})+\frac{\alpha_P\gamma}{2}(r(t)-\pi_t(\hat{Q}))\\
    +\frac{\alpha_I\gamma^2}{4}\int^t dt'(r(t')-\pi_{t'}(\hat{Q}))+\alpha_D2\kappa\mathcal{V}_{Y_a,Q}(\pi_t(\hat{Y}_a)-\hat{Y}_a)\Big]\Big)dt\\
    -\sqrt{\frac{\gamma}{2}}d\hat{B}^\dagger-\sqrt{\frac{\gamma}{2}}d\hat{B}-\frac{\alpha_D}{1+\alpha_D}\sqrt{2\kappa}\mathcal{V}_{Y_a,Q}(-id\hat{A}+id\hat{A}^\dagger)\; ,
\end{multline}
\begin{equation}
    d\hat{P}=\left(G\hat{X}_a-\frac{\gamma}{2}\hat{P}\right)dt-i\sqrt{\frac{\gamma}{2}}d\hat{B}^\dagger+i\sqrt{\frac{\gamma}{2}}d\hat{B}
\end{equation}
\begin{equation}
    d\hat{X}_a=-\frac{\kappa}{2}\hat{X}_a dt-\sqrt{\frac{\kappa}{2}}d\hat{A}^\dagger-\sqrt{\frac{\kappa}{2}}d\hat{A}
\end{equation}
\begin{equation}
    d\hat{Y}_a=\left(G\hat{Q}-\frac{\kappa}{2}\hat{Y}_a\right)dt-i\sqrt{\frac{\kappa}{2}}d\hat{A}^\dagger+i\sqrt{\frac{\kappa}{2}}d\hat{A}\; ,
\end{equation}
\end{subequations}
where the feedback only affects the $\hat{Q}$ quadrature.

\subsection{Equations for filtered estimates}
The feedback Hamiltonian only modifies the BKS equation for the $\hat{Q}$ quadrature to the form:
\begin{multline}\label{eq:Q_new}
    d\pi_t(\hat{Q})=\frac{1}{1+\alpha_D}\Bigg[\Big(-\frac{\gamma}{2}\pi_t(\hat{Q})+\frac{\alpha_P\gamma}{2}(r(t)-\pi_t(\hat{Q}))\\
    +\frac{\alpha_I\gamma^2}{4}\int^t dt'(r(t')-\pi_{t'}(\hat{Q}))\Big)dt+\sqrt{2\kappa}\mathcal{V}_{Y_a,Q}dI\Bigg]\; .
\end{multline}

The ensemble-averaged filtered estimate satisfies the same equation as the expectation value of the corresponding operator, so we may use the estimates to obtain the average time evolution of the quadratures. Averaging Eqs.~\eqref{eq:pi} and~\eqref{eq:Q_new}, we obtain the dynamical equations for the system's quadratures
\begin{subequations}\label{eq:linear}
\begin{multline}
    \frac{d}{dt}\langle\pi_t(\hat{Q})\rangle=\frac{1}{1+\alpha_D}\Big(-\frac{\gamma}{2}\langle\pi_t(\hat{Q})\rangle+\frac{\alpha_P\gamma}{2}(r(t)-\langle\pi_t(\hat{Q})\rangle)\\
    +\frac{\alpha_I\gamma^2}{4}\int^t dt'(r(t')-\langle\pi_{t'}(\hat{Q})\rangle)\Big)
\end{multline}
\begin{equation}
    \frac{d}{dt}\langle\pi_t(\hat{P})\rangle=G\langle\pi_t(\hat{X}_a)\rangle-\frac{\gamma}{2}\langle\pi_t(\hat{P})\rangle
\end{equation}
\begin{equation}
    \frac{d}{dt}\langle\pi_t(\hat{X}_a)\rangle=-\frac{\kappa}{2}\langle\pi_t(\hat{X}_a)\rangle
\end{equation}
\begin{equation}
    \frac{d}{dt}\langle\pi_t(\hat{Y}_a)\rangle=G\langle\pi_t(\hat{Q})\rangle-\frac{\kappa}{2}\langle\pi_t(\hat{Y}_a)\rangle\; .
\end{equation}
\end{subequations}

The average of the quadratic estimate terms may be derived from the linear terms as
\begin{equation}
    d\langle\pi_t(\hat{X})\pi_t(\hat{Y})\rangle=\langle d\pi_t(\hat{X})\pi_t(\hat{Y})\rangle+\langle\pi_t(\hat{X})d\pi_t(\hat{Y})\rangle+\langle d\pi_t(\hat{X})d\pi_t(\hat{Y})\rangle\; ,
\end{equation}
where the last term is the Itô correction.

The quadratic terms relevant for the calculation of the variances of the mechanical quadratures are the following:
\begin{subequations}\label{eq:quadratic}
\begin{multline}\label{eq:Q2}
    \frac{d}{dt}\langle\pi_t(\hat{Q})^2\rangle=\frac{2}{1+\alpha_D}\Big(-\frac{\gamma}{2}\langle\pi_t(\hat{Q})^2\rangle+\frac{\alpha_P\gamma}{2}(r(t)\langle\pi_t(\hat{Q})\rangle-\langle\pi_t(\hat{Q})^2\rangle)\\
    +\frac{\alpha_I\gamma^2}{4}\int^t dt'(r(t')\langle\pi_t(\hat{Q})\rangle-\langle\pi_t(\hat{Q})\pi_{t'}(\hat{Q})\rangle)\Big)+\frac{2\kappa\mathcal{V}^2_{Y_a,Q}}{(1+\alpha_D)^2}
\end{multline}
\begin{equation}
    \frac{d}{dt}\langle\pi_t(\hat{P})^2\rangle=2G\langle\pi_t(\hat{X}_a)\pi_t(\hat{P})\rangle-\gamma\langle\pi_t(\hat{P})^2\rangle+2\kappa\mathcal{V}_{Y_a,P}^2
\end{equation}
\begin{multline}
    \frac{d}{dt}\langle\pi_t(\hat{X_a})\pi_t(\hat{P})\rangle=G\langle\pi_t(\hat{X}_a)^2\rangle\\
    -\left(\frac{\kappa}{2}+\frac{\gamma}{2}\right)\langle\pi_t(\hat{X_a})\pi_t(\hat{P})\rangle+2\kappa\mathcal{V}_{X_a,Y_a}\mathcal{V}_{Y_a,P}
\end{multline}
\begin{equation}
    \frac{d}{dt}\langle\pi_t(\hat{X}_a)^2\rangle=-\kappa\langle\pi_t(\hat{X}_a)^2\rangle+2\kappa\mathcal{V}_{X_a,Y_a}^2
\end{equation}
\end{subequations}
The integral feedback term in Eq.~\eqref{eq:Q2}, $\sigma\equiv\int^t dt'\langle\pi_t(\hat{Q})\pi_{t'}(\hat{Q})\rangle$, involves the average of estimators at different times. The differential equation for $\sigma$ may be obtained by applying the Leibniz integral rule twice:
\begin{widetext}
\begin{multline}\label{eq:sigma}
    \frac{d^2\sigma}{dt^2}=\frac{1}{1+\alpha_D}\left[\left(\frac{\alpha_P\gamma}{2}r(t)+\frac{\alpha_I\gamma^2}{4}\int^t dt' r(t')\right)\langle\pi_t(\hat{Q})\rangle+\left(\frac{\alpha_P\gamma}{2}\frac{dr(t)}{dt}+\frac{\alpha_I\gamma^2}{4}r(t)\right)\int^t dt'\langle\pi_{t'}(\hat{Q})\rangle\right]\\
    -\frac{1}{1+\alpha_D}\frac{(1+\alpha_P)\gamma}{2}\frac{d\sigma}{dt}-\frac{2}{1+\alpha_D}\frac{\alpha_I\gamma^2}{4}\sigma+\frac{d\langle\pi_t(\hat{Q})^2\rangle}{dt}\; .
\end{multline}
\end{widetext}

\subsection{Equations for the filtered estimate covariances}\label{sec:covariances}
The filtered estimate covariances, defined as $\mathcal{V}_{\hat{X},\hat{Y}}=\pi_t(\{\hat{X},\hat{Y}\}/2)-\pi_t(\hat{X})\pi_t(\hat{Y})$ are deterministic functions satisfying the coupled equations
\begin{subequations}\label{eq:covariances}
    \begin{equation}
        \dot{\mathcal{V}}_{Q}=-\gamma\left(\mathcal{V}_{Q}-(n_\mathrm{th}+1/2)\right)-2\kappa\mathcal{V}_{Y_a,Q}^2
    \end{equation}
    \begin{equation}
        \dot{\mathcal{V}}_{Q,P}=G\mathcal{V}_{X_a,Q}-\gamma\mathcal{V}_{Q,P}-2\kappa\mathcal{V}_{Y_a,Q}\mathcal{V}_{Y_a,P}
    \end{equation}
    \begin{equation}
        \dot{\mathcal{V}}_{P}=2G\mathcal{V}_{X_a,P}-\gamma\left(\mathcal{V}_{P}-(n_\mathrm{th}+1/2)\right)-2\kappa\mathcal{V}_{Y_a,P}^2
    \end{equation}
    \begin{equation}
        \dot{\mathcal{V}}_{X_a}=-\kappa\left(\mathcal{V}_{X_a}-\frac{1}{2}\right)-2\kappa\mathcal{V}_{X_a,Y_a}^2
    \end{equation}
    \begin{equation}
        \dot{\mathcal{V}}_{X_a,Y_a}=G\mathcal{V}_{X_a,Q}-\kappa\mathcal{V}_{X_a,Y_a}-2\kappa\mathcal{V}_{X_a,Y_a}\left(\mathcal{V}_{Y_a}-\frac{1}{2}\right)
    \end{equation}
    \begin{equation}
        \dot{\mathcal{V}}_{Y_a}=2G\mathcal{V}_{Y_a,Q}-\kappa\left(\mathcal{V}_{Y_a}-\frac{1}{2}\right)-2\kappa\left(\mathcal{V}_{Y_a}-\frac{1}{2}\right)^2
    \end{equation}
    \begin{equation}
        \dot{\mathcal{V}}_{X_a,Q}=-\left(\frac{\kappa}{2}+\frac{\gamma}{2}\right)\mathcal{V}_{X_a,Q}-2\kappa\mathcal{V}_{X_a,Y_a}\mathcal{V}_{Y_a,Q}
    \end{equation}
    \begin{equation}
        \dot{\mathcal{V}}_{X_a,P}=G\mathcal{V}_{X_a}-\left(\frac{\kappa}{2}+\frac{\gamma}{2}\right)\mathcal{V}_{X_a,P}-2\kappa\mathcal{V}_{X_a,Y_a}\mathcal{V}_{Y_a,P}
    \end{equation}
    \begin{equation}
        \dot{\mathcal{V}}_{Y_a,Q}=G\mathcal{V}_{Q}-\left(\frac{\kappa}{2}+\frac{\gamma}{2}\right)\mathcal{V}_{Y_a,Q}-2\kappa\left(\mathcal{V}_{Y_a}-\frac{1}{2}\right)\mathcal{V}_{Y_a,Q}
    \end{equation}
    \begin{equation}
        \dot{\mathcal{V}}_{Y_a,P}=G(\mathcal{V}_{X_a,Y_a}+\mathcal{V}_{Q,P})-\left(\frac{\kappa}{2}+\frac{\gamma}{2}\right)\mathcal{V}_{Y_a,P}-2\kappa\mathcal{V}_{Y_a,P}\left(\mathcal{V}_{Y_a}-\frac{1}{2}\right)\; ,
    \end{equation}
\end{subequations}
with the initial values of the quadrature variances given by the thermal fluctuations and the other covariances are zero. From equations~(\ref{eq:linear}-\ref{eq:covariances}), we may derive the conditional and unconditional variances of the system operators.

\section{Tuning of the PID parameters}\label{sec:tuning}
In this section, we explain how the PID parameters may be tuned to satisfy speed and stability specifications, for example, to initialize a mechanical quadrature in a force-sensing experiment within the desired time frame. We use frequency domain analysis to describe the dynamics of the system. The position of the zeros and poles in the transfer function of the system determine the dynamics of the system, so the PID parameters may be tuned to transform the denominator of the transfer function into some desired second-order system of the form $s^2+2\varsigma\omega_n s+\omega_n^2$.

As an example, we consider an overshooting of $R=5\%$ and a stabilization time of $T_p=2/\gamma$. In a \textit{standard} second-order system with no zeros, these specifications depend on the poles of the transfer function as $R=e^{-\pi\varsigma/\sqrt{1-\varsigma^2}}$ and $T_p=4/(\varsigma\omega_n)$. Plugging the denominator of the transfer function~\eqref{eq:G} into the formulas, the corresponding PID parameters are the following: $\alpha_P=7$, $\alpha_I=33.6$ and $\alpha_D=0$. The time-response of the quadrature is shown in the inset of Fig.~\ref{fig:control_quadrature}(a). The system does not satisfy the required specifications due to the extra dynamics introduced by the zero of the transfer function. This may be overcome by weighting the setpoint in the proportional feedback $\hat{H}_\mathrm{P}\propto \mu r(t)-\pi_t(\hat{Q})$. The scaling does not affect squeezing, as it only affects the reference signal. For $\mu=0$, the P term in the numerator vanishes, and the system becomes a \textit{standard} second-order system, satisfying the specifications accordingly.

\section{Measurement of a weak force}\label{sec:force}
The measurement of a mechanical quadrature may be used for the detection of weak external forces. Deviations from the predicted evolution of the quadratures provide information about the forces acting on the system. The mechanical quadratures $\hat{Q}$ and $\hat{P}$ describe the in-phase and out-of-phase components of the motion at frequency $\omega_m$:
\begin{equation}
    \hat{x}(t)=\hat{Q}(t)\cos(\omega_m t)+\hat{P}(t)\sin(\omega_m t)
\end{equation}
An external sinusoidal force
\begin{equation}\label{eq:F}
    F_\mathrm{ext}(t)=f_\mathrm{ext}\sin{(\omega t+\varphi)}=f_1\sin{(\omega t)}+f_2\cos{(\omega t)}
\end{equation}
which is near-resonant with the mechanical frequency $|\omega-\omega_m|\lesssim\gamma$, induces a steady-state displacement of the quadratures. If the displacement is greater than the measured quadrature uncertainty, the measurement may be used to detect the force.

The measurement of one quadrature induces back-action on the opposite quadrature, which makes simultaneous accurate measurement of $\hat{Q}$ and $\hat{P}$, and therefore $f_1$ and $f_2$, impossible. However, one may use a second resonator coupled to the same force to squeeze and measure the second quadrature, providing both components of the force $F(t)$.

The Hamiltonian corresponding to the external force~\eqref{eq:F} in the rotating frame, within the rotating wave approximation, takes the form
\begin{equation}
    \hat{H}_\mathrm{ext}=-\hbar\frac{f_1}{2}\hat{P}-\hbar\frac{f_2}{2}\hat{Q}\; .
\end{equation}
The external force modifies the time evolution of the mechanical quadratures [Eq.~\eqref{eq:pi}] as
\begin{subequations}
\begin{equation}
    \frac{d}{dt}\langle \hat{Q}\rangle=-\frac{f_1}{2}-\frac{\gamma}{2}\langle\hat{Q}\rangle
\end{equation}
\begin{equation}
    \frac{d}{dt}\langle\hat{P}\rangle=\frac{f_2}{2}+G\langle\hat{X}_a\rangle-\frac{\gamma}{2}\langle\hat{P}\rangle
\end{equation}
\end{subequations}
which leads to a steady-state displacement of the quadratures of $\langle\hat{Q}\rangle=-f_1/\gamma$ and $\langle\hat{P}\rangle=f_2/\gamma$. The system will evolve into the steady state over a characteristic time of the order of $\gamma^{-1}$. PID feedback may be used to shorten the system's response time to the external force, as described in the main text. As a result, this improves the sensitivity of the measurement to the force variations on time scales much smaller than $\gamma^{-1}$.

\section{Treatment of an unknown external force}\label{sec:f_ext}
In this section, we show how to implement an unknown external force into the Kalman filter model. We consider a force of the form $F_\mathrm{ext}(t)=f_\mathrm{ext}(t)\sin{(\omega_m t)}$, which we want to determine from a $\hat{Q}$ quadrature measurement. We cannot include the Hamiltonian term corresponding to the external force
\begin{equation}
    \hat{H}_\mathrm{ext}=-\hbar\frac{f_\mathrm{ext}(t)}{2}\hat{P}\; .
\end{equation}
directly into the BKS equations~\eqref{eq:BKS}, since the function $f_\mathrm{ext}(t)$ is unknown. Instead, we augment the Kalman filter to include the
estimate of the classical variable $f_\mathrm{ext}(t)$. We can treat the estimate $\pi_t(f_\mathrm{ext})$ within the quantum Kalman filter formalism by considering an operator proportional to the identity: $\hat{f}_\mathrm{ext}=f_\mathrm{ext}\mathds{1}$. If we assume that the external force remains roughly constant within the characteristic mechanical time scale $\gamma^{-1}$, $\partial f_\mathrm{ext}/\partial t \approx 0$ so the system dynamics for $\pi_t(f_\mathrm{ext})$ will be trivial. The innovations process $dI$~\eqref{eq:dI} will update the value of $\pi_t(f_\mathrm{ext})$ according to
\begin{equation}\label{eq:f}
    d\pi_t(f_\mathrm{ext})=\sqrt{2\kappa}\mathcal{V}_{Y_a,f}dI\; .
\end{equation}
The greater the mismatch between the output field and our prediction, the greater $dI$ will be, so that $\pi_t(f_\mathrm{ext})$ approaches the real value of $f_\mathrm{ext}$ in the stationary regime.

Including the estimate of the force into the Kalman filter model, the BKS equation for quadrature $\hat{Q}$ is modified as
\begin{multline}\label{eq:Q_force}
    d\pi_t(\hat{Q})=\frac{1}{1+\alpha_D}\Big[\Big(-\frac{1}{2}\pi_t(f_\mathrm{ext})-\frac{\gamma}{2}\pi_t(\hat{Q})+\frac{\alpha_P\gamma}{2}(r(t)-\pi_t(\hat{Q}))\\
    +\frac{\alpha_I\gamma^2}{4}\int^t dt'(r(t')-\pi_{t'}(\hat{Q}))\Big)dt+\sqrt{2\kappa}\mathcal{V}_{Y_a,Q}dI\Big]
\end{multline}
Due to the imperfect modeling of the external force dynamics, the ensemble-averaged filtered estimate of the quadratures will differ from their true value $\langle\pi_t(\hat{Q})\rangle\neq\langle\hat{Q}\rangle$, so the innovation process is no longer a martingale $\langle dI\rangle \neq 0$. From Eq.~\eqref{eq:dI}, we obtain
\begin{equation}
    \left\langle\frac{dI}{dt}\right\rangle=\sqrt{2\kappa}(\langle\hat{Y}_a\rangle-\langle\pi_t(\hat{Y}_a)\rangle)\; .
\end{equation}

The equations of motion for the real quadratures are given by the Heisenberg-Langevin equations~\eqref{eq:linear_real}. The equation for the $\hat{Q}$ quadrature is modified by the external force as
\begin{widetext}
\begin{multline}\label{eq:linear_real2}
    \frac{d}{dt}\langle\hat{Q}\rangle=-\frac{1}{2}f_\mathrm{ext}(t)-\frac{\gamma}{2}\langle\hat{Q}\rangle+\frac{1}{1+\alpha_D}\Big[\alpha_D\frac{1}{2}\pi_t(f_\mathrm{ext})+\alpha_D\frac{\gamma}{2}\langle\pi_t(\hat{Q})\rangle+\frac{\alpha_P\gamma}{2}(r(t)-\langle\pi_t(\hat{Q})\rangle)\\
    +\frac{\alpha_I\gamma^2}{4}\int^t dt'(r(t')-\langle\pi_{t'}(\hat{Q})\rangle)+\alpha_D2\kappa\mathcal{V}_{Y_a,Q}(\langle\pi_t(\hat{Y}_a)\rangle-\langle\hat{Y}_a\rangle)\Big]\; .
\end{multline}
\end{widetext}
In the main text, we solve the coupled system of equations for the averages of the operators~\eqref{eq:linear_real} and~\eqref{eq:linear_real2}, their estimates~\eqref{eq:linear} and~\eqref{eq:Q_force}, and the estimate of the external force~\eqref{eq:f}, for different values of the PID parameters.

\subsection{Equations for the filtered estimate covariances}\label{sec:covariances2}
The addition of the estimate of the external force into the Kalman filter increases the number of covariances to be determined. The coupled system of equations for the covariances is:
\begin{subequations}\label{eq:covariances2}
    \begin{equation}
        \dot{\mathcal{V}}_{Q}=-\gamma\left(\mathcal{V}_{Q}-(n_\mathrm{th}+1/2)\right)-2\kappa\mathcal{V}_{Y_a,Q}^2-\mathcal{V}_{Q,f}
    \end{equation}
    \begin{equation}
        \dot{\mathcal{V}}_{Q,P}=G\mathcal{V}_{X_a,Q}-\gamma\mathcal{V}_{Q,P}-2\kappa\mathcal{V}_{Y_a,Q}\mathcal{V}_{Y_a,P}-\frac{1}{2}\mathcal{V}_{P,f}
    \end{equation}
    \begin{equation}
        \dot{\mathcal{V}}_{P}=2G\mathcal{V}_{X_a,P}-\gamma\left(\mathcal{V}_{P}-(n_\mathrm{th}+1/2)\right)-2\kappa\mathcal{V}_{Y_a,P}^2
    \end{equation}
    \begin{equation}
        \dot{\mathcal{V}}_{X_a}=-\kappa\left(\mathcal{V}_{X_a}-\frac{1}{2}\right)-2\kappa\mathcal{V}_{X_a,Y_a}^2
    \end{equation}
    \begin{equation}
        \dot{\mathcal{V}}_{X_a,Y_a}=G\mathcal{V}_{X_a,Q}-\kappa\mathcal{V}_{X_a,Y_a}-2\kappa\mathcal{V}_{X_a,Y_a}\left(\mathcal{V}_{Y_a}-\frac{1}{2}\right)
    \end{equation}
    \begin{equation}
        \dot{\mathcal{V}}_{Y_a}=2G\mathcal{V}_{Y_a,Q}-\kappa\left(\mathcal{V}_{Y_a}-\frac{1}{2}\right)-2\kappa\left(\mathcal{V}_{Y_a}-\frac{1}{2}\right)^2
    \end{equation}
    \begin{equation}
        \dot{\mathcal{V}}_{X_a,Q}=-\left(\frac{\kappa}{2}+\frac{\gamma}{2}\right)\mathcal{V}_{X_a,Q}-2\kappa\mathcal{V}_{X_a,Y_a}\mathcal{V}_{Y_a,Q}-\frac{1}{2}\mathcal{V}_{X_a,f}
    \end{equation}
    \begin{equation}
        \dot{\mathcal{V}}_{X_a,P}=G\mathcal{V}_{X_a}-\left(\frac{\kappa}{2}+\frac{\gamma}{2}\right)\mathcal{V}_{X_a,P}-2\kappa\mathcal{V}_{X_a,Y_a}\mathcal{V}_{Y_a,P}
    \end{equation}
    \begin{equation}
        \dot{\mathcal{V}}_{Y_a,Q}=G\mathcal{V}_{Q}-\left(\frac{\kappa}{2}+\frac{\gamma}{2}\right)\mathcal{V}_{Y_a,Q}-2\kappa\left(\mathcal{V}_{Y_a}-\frac{1}{2}\right)\mathcal{V}_{Y_a,Q}-\frac{1}{2}\mathcal{V}_{Y_a,f}
    \end{equation}
    \begin{equation}
        \dot{\mathcal{V}}_{Y_a,P}=G(\mathcal{V}_{X_a,Y_a}+\mathcal{V}_{Q,P})-\left(\frac{\kappa}{2}+\frac{\gamma}{2}\right)\mathcal{V}_{Y_a,P}-2\kappa\mathcal{V}_{Y_a,P}\left(\mathcal{V}_{Y_a}-\frac{1}{2}\right)
    \end{equation}
    \begin{equation}
        \dot{\mathcal{V}}_{Q,f}=-\frac{1}{2}\mathcal{V}_f-\frac{\gamma}{2}\mathcal{V}_{Q,f}-2\kappa\mathcal{V}_{Y_a,Q}\mathcal{V}_{Y_a,f}\; .
    \end{equation}
    \begin{equation}
        \dot{\mathcal{V}}_{P,f}=G\mathcal{V}_{X_a,f}-\frac{\gamma}{2}\mathcal{V}_{P,f}-2\kappa\mathcal{V}_{Y_a,P}\mathcal{V}_{Y_a,f}
    \end{equation}
    \begin{equation}
        \dot{\mathcal{V}}_{X_a,f}=-\frac{\kappa}{2}\mathcal{V}_{X_a,f}-2\kappa\mathcal{V}_{X_a,Y_a}\mathcal{V}_{Y_a,f}
    \end{equation}
    \begin{equation}
        \dot{\mathcal{V}}_{Y_a,f}=G\mathcal{V}_{Q,f}-\frac{\kappa}{2}\mathcal{V}_{Y_a,f}-2\kappa\left(\mathcal{V}_{Y_a}-\frac{1}{2}\right)\mathcal{V}_{Y_a,f}
    \end{equation}
    \begin{equation}
        \dot{\mathcal{V}}_{f}=-2\kappa\mathcal{V}_{Y_a,f}^2\; .
    \end{equation}
\end{subequations}
For the initial value of the force variance $\mathcal{V}_f$, one should select a value that leads to a fast enough update of the force estimate, without making the system unstable. An appropriate value is a variance the order of the expected magnitude of the force squared $\mathcal{V}_f(0)\lesssim f_\mathrm{ext}^2$.

%

\end{document}